\documentclass[fleqn,usenatbib,onecolumn]{mnras}

\usepackage{newtxtext,newtxmath}

\usepackage[T1]{fontenc}
\usepackage{ae,aecompl}

\usepackage{graphicx}	
\usepackage{amsmath}	
\usepackage{amssymb}	
\usepackage{lineno}
\usepackage{psfig}
 
\title[Globular cluster population of J07173724]{Globular cluster population of the {\it HST} frontier fields galaxy J07173724+3744224}

\author[N. Carlson et al.]{
Nathan L. Carlson,$^{1}$
Madina R. Sultanova,$^{1}$
Sandanuwan P. Kalawila Vithanage,$^{1}$
\newauthor
Wayne A. Barkhouse,$^{1}$\thanks{E-mail: wayne.barkhouse@und.edu}
Gihan L. Ipita Kaduwa Gamage,$^{1}$
Cody M. Rude$^{1,2}$
\newauthor
and Omar L{\'o}pez-Cruz$^{3,4}$
\\
$^{1}$Department of Physics and Astrophysics, University of North Dakota, Grand Forks, ND 58202-7129, USA\\
$^{2}$Massachusetts Institute of Technology, Haystack Observatory, Westford, MA 01886-1299, USA\\
$^{3}$Oliver L. Benediktson Endowed Chair in Astrophysics, Department of Physics and Astrophysics, University of 
North Dakota,\\ Grand Forks, ND 58202-7129, USA\\
$^{4}$INAOE, Tonantzintla, 72840, Puebla, Mexico}

\date{Accepted XXX. Received YYY; in original form ZZZ}

\pubyear{2017}

\begin{document}
\label{firstpage}
\pagerange{\pageref{firstpage}--\pageref{lastpage}}
\maketitle

\begin{abstract}
We present the first measurement of the globular cluster population surrounding the elliptical galaxy J07173724+3744224 ($z=0.1546$). This galaxy is located 
in the foreground in the field-of-view of the {\it Hubble Space Telescope} ({\it HST}) Frontier Fields observations of galaxy cluster MACS J0717.5+3745 
($z=0.5458$).
Based on deep {\it HST} ACS {\it F435W}, 
{\it F606W}, and {\it F814W} images, we find a total globular cluster population of $N_{tot}=3441\pm 1416$. Applying the appropriate extinction correction and filter 
transformation from ACS {\it F814W} to the Johnson {\it V}-band, we determine that the host galaxy has an absolute magnitude of $M_{V}=-22.2$. The 
specific frequency was found to be $S_{N}=4.5 \pm 1.8$. The radial profile of the globular cluster system was best fit using a powerlaw of the form 
$\sigma\sim R^{-0.6}$, with the globular cluster population found to be more extended than the halo light of the host galaxy ($\sigma_{halo}\sim R^{-1.7}$). 
The {\it F435W}$-${\it F814W} colour distribution suggests a bimodal population, with red globular clusters 
$1-3\times$ more abundant than blue clusters. These results are consistent with the host elliptical galaxy J07173724+3744224 having formed its red metal-rich GCs 
in situ, with the blue metal-poor globular clusters accreted from low-mass galaxies. 

\end{abstract}




\section{Introduction}

The globular cluster (GC) population of galaxies has been used for many years to investigate the formation of both isolated and cluster
galaxies \citep[e.g.][]{Harris91,West93,West95,Cote98,Barkhouse01,Goudfrooij03,Harris09,Harris13}. It has become apparent that massive
galaxies in rich clusters contain a large number of globular clusters (GCs) \citep[$\la 1\times 10^{6}$;][]{Harris16,Lee16}, while low-mass
dwarf galaxies in poor environments, in general, have few GCs \citep[e.g.][]{Lane13}. To compare the number of GCs per galaxy, \cite{Harris81} introduced
the specific frequency, defined as the number of GCs per unit galaxy luminosity, normalised to $M_{V}=-15$. Specific frequency,
$S_{N}=N_{tot}10^{0.4(M_{V}+15)}$, has been shown to vary from $S_{N}\sim 1$ for isolated spiral galaxies, to as high as $S_{N}\sim 20$ for brightest
cluster galaxies at the centre of rich clusters \citep[e.g.][]{Harris16}. It is interesting to note that a recent study by \cite{vandokkum17} 
show that several ultra diffuse galaxies in the Coma cluster have extreme values of $S_{N}\sim 100$.

A long standing problem in the study of GCs is to find which factor(s) has the greatest influence in determining the makeup of the population
of GCs for a particular galaxy. GC formation scenarios based on mergers \citep[e.g.][]{Ashman92,Zepf93}, accretion \citep[e.g.][]{Cote98,Cote00}, and 
in situ models \citep{Forbes97} have been proposed, but no model has been successful in explaining observations of GCs for galaxies with a wide range
in mass populating a variety of environments \citep{Peng06,Harris13}.

During the past several decades many studies have investigated the colour distribution of GCs in an attempt to test various GC formation scenarios 
\citep[e.g.][]{Ashman94,Forbes97,Cote98,Brodie06,Harris09,Faifer11,Tonini13,Harris17}. Early studies of the Milky Way GC population provided 
evidence for a bimodal distribution of colours \citep[e.g.][]{Zinn85}. For GC systems associated with other galaxies, investigators uncovered multimodal 
colour distributions, with bimodality being the most common \citep[e.g.][]{Brodie06,Jordan07,Muratov10,Blom12,Escudero15,Harris17}. The presence of `blue' and `red' stellar 
populations suggests differences in metallicity, age, or some combination thereof \citep[i.e. the well-known age-metallicity degeneracy;][]{Worthey94}. 
Recent studies have shown that age differences are too small to account for a bimodal colour distribution and that GCs have ages $>10$ Gyr, thus a 
metallicity effect is expected to dominate the colour spread in GC systems \citep{Puzia05,Strader05,Brodie06,Norris08,Forbes15,Usher15}.   

To increase the diversity of environments in which the GC system of elliptical galaxies have been measured, we studied the GC population of the
elliptical galaxy J07173724+3744224 (hereafter J07173724). This galaxy is in the foreground of the galaxy cluster MACS J07175+3745 ($z=0.5458$),
which is part of the {\it Hubble Space Telescope} ({\it HST}) Frontier Fields observation campaign \citep{Koekemoer14,Lotz17}. The redshift of
J07173724 ($z=0.1546\pm 0.0001$) was measured by \cite{Ebeling14} using the LRIS instrument on Keck-I, but was not included in their published
catalogue (H. Ebeling, private communications). Adopting $\Omega_{\mbox{m}}= 0.3$, $\Omega_{\Lambda}= 0.7$, and
$\mbox{H}_{0}= 70~\mbox{km}~\mbox{s}^{-1}~\mbox{Mpc}^{-1}$ for this study, J07173724 has a distance modulus of $(m-M)=39.34$ and a luminosity
distance of 737.1 Mpc, which yields a physical scale of $2.68~\mbox{kpc}~\mbox{arcsec}^{-1}$. 

J07173724 ($\alpha_{J2000}=07^{\rmn{h}} 17^{\rmn{m}} 37\fs2$, $\delta_{J2000}=37\degr 44\arcmin 23\farcs 0$) is one of the most distant galaxies to have its GC 
population measured, third behind the brightest cluster galaxy in Abell 2744 
\citep[$z=0.308$;][]{Lee16} and Abell 1689 \citep[$z=0.18$;][]{Alamo13}. We also note that J07173724 is catalogued as a FR-I type \citep{Fanaroff74} 
radio galaxy \citep{Bonafede09,vanWeeren16} with a linear-like feature having a total projected length of $\sim\,240$ kpc, as measured from a combined 2-4 GHz
JVLA image \citep[see fig. 1 from][]{vanWeeren17}. \citet{vanWeeren16} report a {\it Chandra} X-ray source coincident with the photometric centroid of 
J07173724 having an X-ray flux of $(5.88_{-0.14}^{+0.15})\times 10^{-15}$~erg~cm$^{-2}$~s$^{-1}$ ($2-10$ keV; see their table 5). Using our adopted luminosity 
distance, this yields an X-ray luminosity of $(3.82_{-0.09}^{+0.10})\times 10^{41}$~erg~s$^{-1}$. The lens model from \citet{Diego15} for the MACS J07175+3745 
field yields a mass for J07173724 of $(8.88\pm 0.14)\times 10^{11}$~M$_{\sun}$ (J. Diego, private communications). 

In this paper observations and data reductions are described in Section 2. In Section 3 we present our results regarding the GC colour
distribution, luminosity function, radial profile, and specific frequency. Our discussion is given in Section 4, and our conclusions are
outlined in Section 5.

\section{Observations and data reductions}
\label{Observations}

Observations of galaxy J07173724 consists of {\it HST} ACS 30-mas ($0.03~\mbox{arcsec}~\mbox{pixel}^{-1}$) self-calibrated epoch 1.0 images of galaxy cluster
MACS J07175+3745, downloaded from the Mikulski Archive for Space Telescopes (MAST) as part of the Frontier Fields program \citep{Lotz17}. We elected to use 
{\it F435W}, {\it F606W}, and {\it F814W} images due to their depth, with effective co-add exposure times of 47746 s ({\it F435W}), 27015 s ({\it F606W}), 
and 114591 s ({\it F814W}), and the appearance in all three filters of an excess of starlike objects centred on J07173724. Extinction corrections of 
0.277 mag ({\it F435W}), 0.190 mag ({\it F606W}), and 0.117 mag ({\it F814W}) were taken from \cite{Schlafly11}. Extinction corrections were applied to the 
AB magnitude system zero-points, which are available at \url{https://archive.stsci.edu/prepds/frontier/}.

In addition to images of MACS J07175+3745, we also acquired {\it F435W}, {\it F606W}, and {\it F814W} 30-mas self-calibrated {\it HST} ACS epoch 1.0 images 
of the nearby MACS J07175+3745 parallel field. The centre of the parallel field is $\sim 6.7$ arcmin from the photometric centroid of J07173724, and is 
located in a northwest direction from the galaxy. The extinction corrections from \cite{Schlafly11} for the parallel field are 0.275 mag ({\it F435W}), 
0.188 mag ({\it F606W}), and 0.116 mag ({\it F814W}), and were applied to the AB magnitude zero-points. The co-add total exposure times for the three filters 
are 70636/39816/109750 seconds for the {\it F435W}/{\it F606W}/{\it F814W} filters. The parallel field has a comparable depth to the galaxy images, and is used 
to sample the background field population to statistically subtract GC-like objects from our galaxy field detections (see Sec \ref{Background} and 
\ref{Backcontamination}).

To maximize the S/N for object detection on the galaxy and parallel fields (see Section \ref{PPP}), we used the task \textsc{imcombine} in 
\textsc{iraf}\footnote{\textsc{iraf} is distributed by the National Optical Astronomy Observatory, which is operated by the Association of Universities 
for Research in Astronomy, Inc., under cooperative agreement with the National Science Foundation.} to sum together all three filter images into two 
separate `combined' galaxy and parallel fields. The combined frames are used to generate object catalogues, while photometry is performed on the 
separate filter images for the galaxy and parallel frames.

\subsection{Removal of host galaxy light}

To perform object detection and photometry of GC candidates, we must remove the light from the host galaxy halo. The galaxy light was removed by first
using the \textsc{ellipse} task in the \textsc{iraf/stsdas} package to fit elliptical isophotes to our 2D images following the method of \cite{Jedrzejewski87}. 
The isophotes were fit to the galaxy light out to a maximum radius of 800 pixels (24 arcsec or 64 kpc). This maximum size was chosen based on the radius at 
which the galaxy light profile, from visual inspection, merged into the background light of the cluster field. The \textsc{stsdas/bmodel} task was then used to 
build a model of the light distribution of the galaxy to a radius of 800 pixels, and this model was subtracted from the parent image. A constant value was 
then added to the subtracted image to restore the background counts.   

The removal of the galaxy light by the \textsc{iraf/stsdas} software revealed the presence of an artefact (possible dust-lane) near the centre of the galaxy. 
To construct a very flat image for object detection on the galaxy frame, we applied a ring median filter to the combined galaxy-subtracted image using the task 
\textsc{rmedian} in \textsc{iraf} (inner radius = 5 pixels, outer radius = 9 pixels). The median filtered image was subtracted from the input image, and the 
background counts were restored to the original values. This process helped us to generate a flatter image for object detection with minimal galaxy residual. 
The median-filtered step was not applied to the individual filter images of the galaxy field (which are used for photometry) since the flux of detected objects 
is altered by the median filtering process. 

\begin{figure*}
\resizebox{0.45\hsize}{!}{\includegraphics{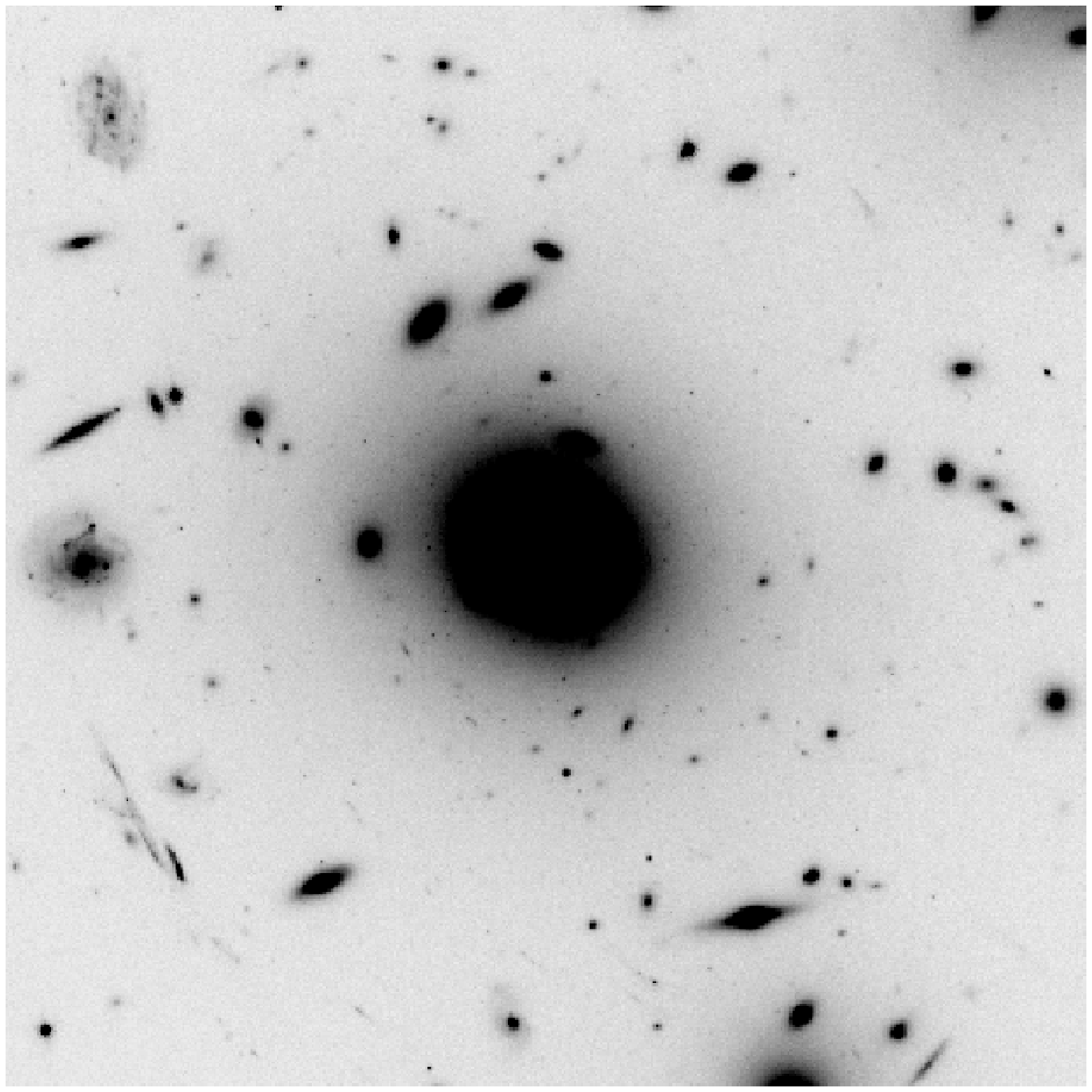}}
\resizebox{0.45\hsize}{!}{\includegraphics{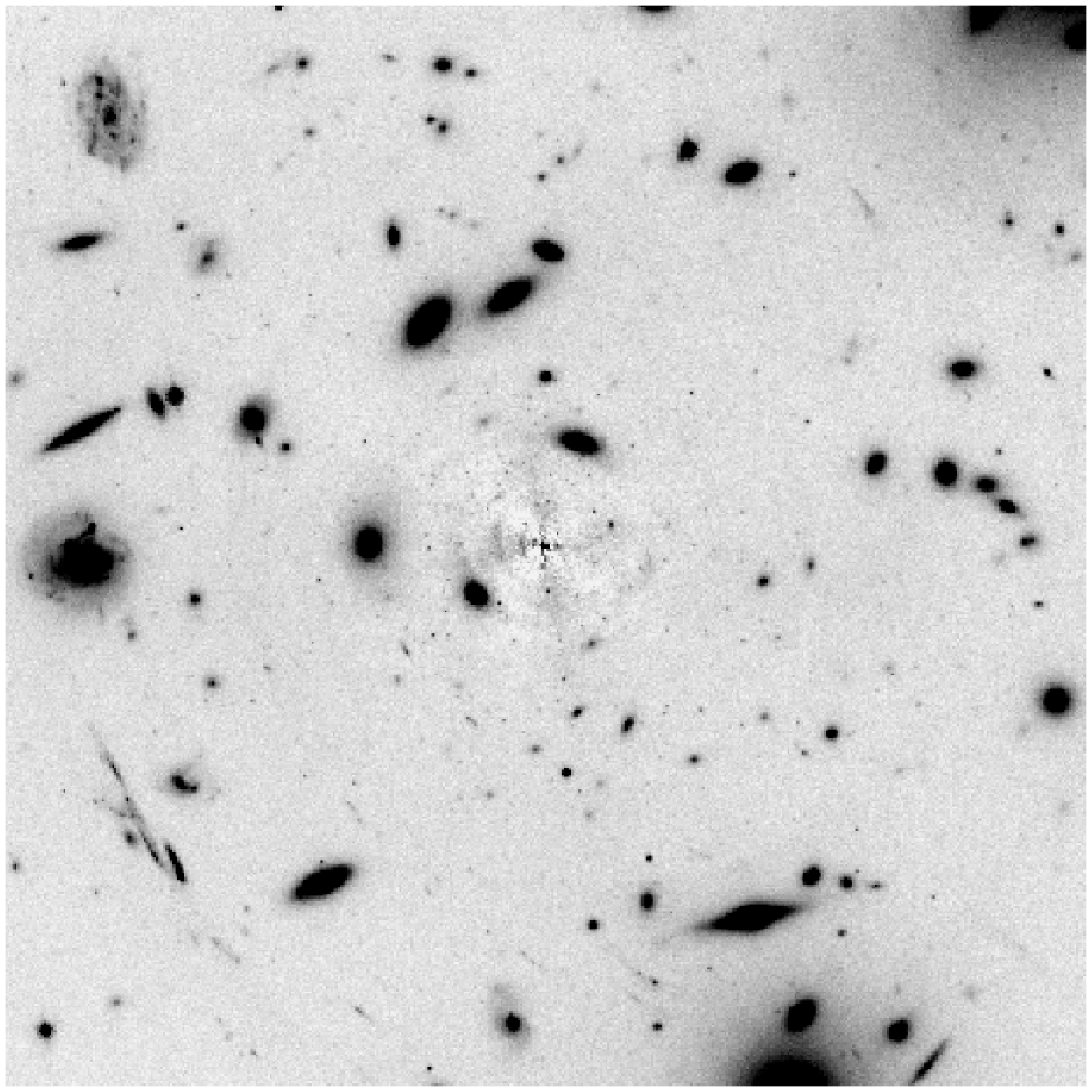}}
\caption{Left: The {\it HST} Frontier Fields ACS {\it F814W} image of J07173724, with a spatial area of $\sim0.7\times 0.7$ arcmin. Right: 
The \textsc{ellipse/bmodel} subtracted image of J07173724, covering the same field-of-view as the left panel. For both images, north is up and 
east is to the left.}
\label{GalaxyRemoval}
\end{figure*}

\begin{figure}
\includegraphics[width=\columnwidth]{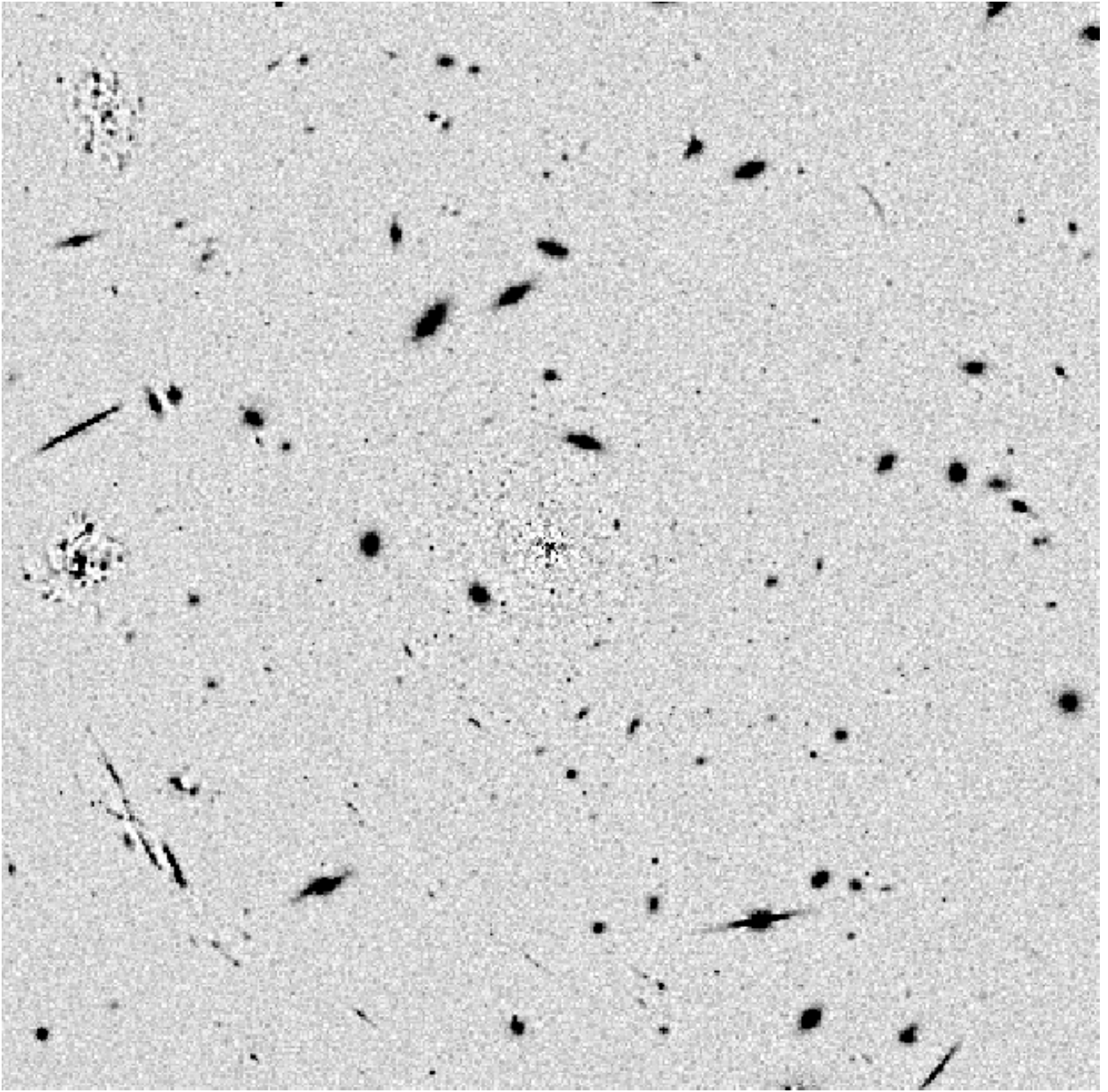}
\caption{The median filtered-subtracted {\it F814W} image of J07173724. The \textsc{ellipse/bmodel} subtracted image depicted in Fig. 1 (right panel) was 
median filtered to produce a flatter image in which to perform object detection.}
\label{GalaxyMedianFilter}
\end{figure}

In Fig. \ref{GalaxyRemoval} we show the results of galaxy subtraction by comparing the {\it F814W} image of J07173724 before and after galaxy light 
removal. The median-filtered combined galaxy image that was used for object detection is shown in Fig. \ref{GalaxyMedianFilter}. The images of the galaxy 
with its halo light removed shows a concentration of starlike objects near the galaxy centroid. These are the globular cluster candidates that we 
seek to detect and measure their properties.

\subsection{Object detection, classification, and photometry}
\label{PPP}

Object detection was performed on the combined galaxy-subtracted, median-filtered, image using the Picture Processing Program  \citep[\textsc{ppp};][]{Yee91}. 
Combining all three filters ({\it F435W}, {\it F606W}, {\it F814W}) for the galaxy field allows us to probe to fainter flux limits for 
object detection than possible using a single filter frame. The master object position file created by \textsc{ppp} using the combined images was used as input to 
the \textsc{iraf} implementation of \textsc{daophot} in order to perform object photometry and classification. Object photometry was done using the 
\textsc{phot} and \textsc{allstar} tasks to measure PSF magnitudes in each separate filter. Part of this process was the construction of a PSF model, 
using \textsc{pstselect} and \textsc{psf}, from a large number of bright, unsaturated, isolated, starlike objects. A separate PSF model, quadratically 
depended on x and y pixel coordinates, was made for each filter, and the resultant model was visually inspected using \textsc{seepsf}.

Aperture corrections for each filter were calculated by comparing PSF magnitudes for a large number of isolated, unsaturated stars, output from \textsc{allstar} 
with \textsc{phot} magnitudes measured using a 0.5 arcsec radius (17 pixel) aperture. The filter-dependent aperture corrections were then extrapolated 
from a 0.5 arcsec radius to infinity based on table 5 from \cite{Sirianni05}. Total aperture corrections were found to be $-0.321\pm 0.008$ mag 
({\it F435W}), $-0.536\pm 0.003$ mag ({\it F606W}), and $-0.607\pm 0.007$ mag ({\it F814W}). 
 
\begin{figure}
\includegraphics[width=\columnwidth]{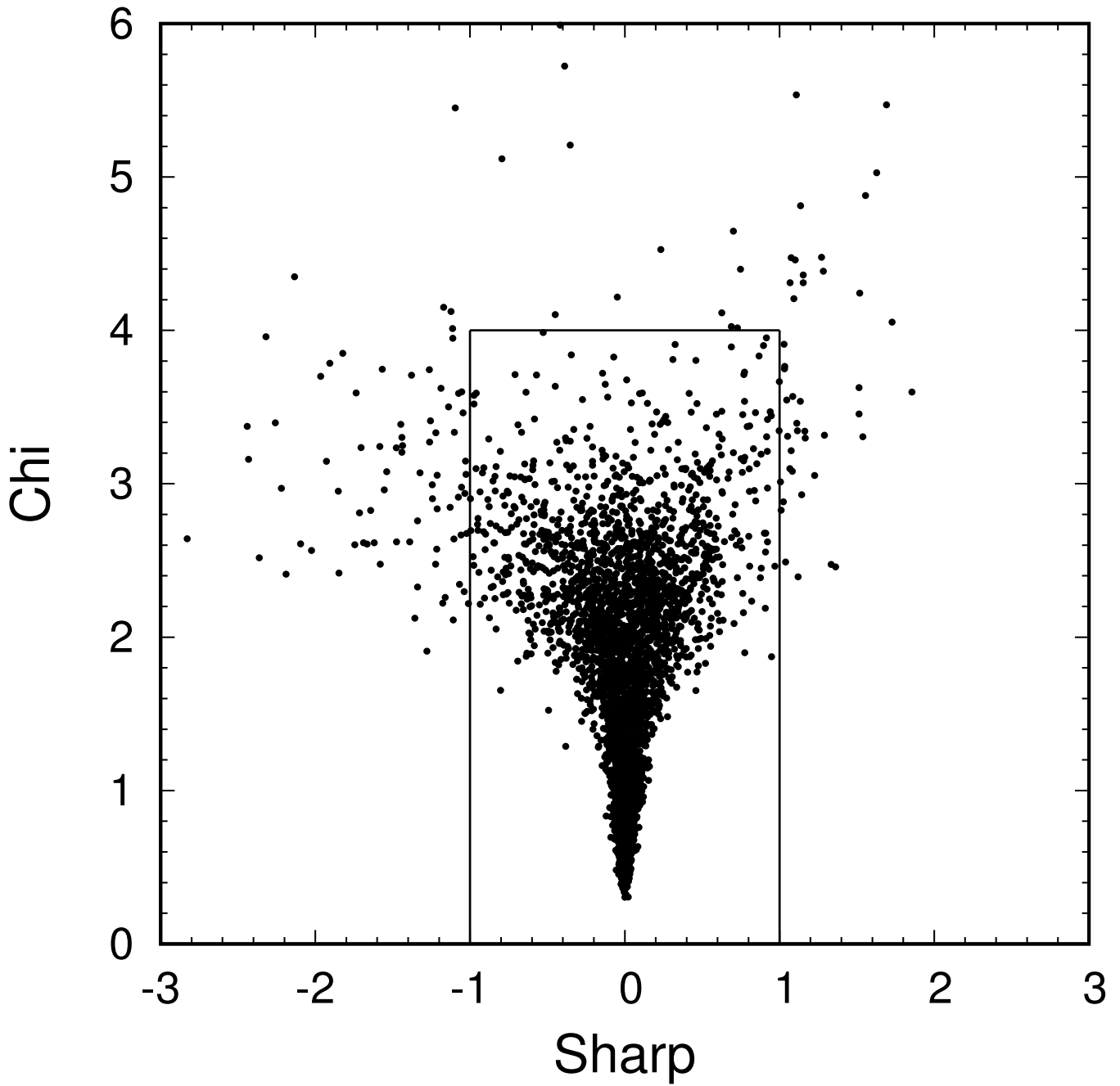}
\caption{Distribution of \textsc{daophot/allstar} {\it sharp} and {\it chi} shape measurements of 4000 artificial stars. Approximately 96 per cent of the starlike objects 
are recovered using $-1< {\it sharp} < +1$ and ${\it chi} < 4$.}   
\label{ShapeParameters}
\end{figure}

GCs associated with J07173724 ($d_{A}=553.0~\mbox{Mpc}$) are unresolved since the average GC \citep[half-radius $\sim 3~\mbox{pc}$;][]{Harris09} at the distance 
of the galaxy is expected to subtend an angle of $\sim 0.2~\mbox{mas}$, which is smaller than the 0.1 arcsec FWHM of the {\it HST} ACS detector. Since GCs 
should appear starlike, we used the \textsc{allstar} shape parameters {\it sharp} and {\it chi} to select GC candidates. The range in the values 
of {\it sharp} and {\it chi} used for GC selection was determined by adding 4000 artificial stars to the {\it F814W} galaxy-subtracted image. The {\it F814W} 
image was used since it has the highest S/N of the GC population of J07173724 (based on visual inspection). Artificial stars were added to 40 separate images 
in groups of 100 (to avoid crowding effects) using the {\it F814W} PSF model and \textsc{addstar}. Artificial stars were assigned random positions away from 
bright objects and image boundary, with {\it F814W} magnitudes randomly selected from 25 to 31 mag. Fig. \ref{ShapeParameters} depicts the range in 
{\it sharp} and {\it chi} values measured for the total sample of 4000 artificial stars. Examination of Fig. \ref{ShapeParameters} shows that a majority 
(96 per cent) of artificial stars are found in a region outlined by $-1< sharp < +1$ and $chi < 4$. Thus we have adopted these ranges in the shape parameters 
for culling GC candidates. In addition to selecting GCs based on shape parameters, we have excluded objects that are $r>470~\mbox{pixels}$ (14 arcsec, 38 kpc) 
away from the centre of J07173724 since the density of starlike objects drops to background levels at this radius. Also, the presence of numerous galaxies in 
the field makes it more problematic to find GC candidates. We also exclude objects detected within $r<20~\mbox{pixels}$ (0.6 arcsec, 1.6 kpc) of the galaxy 
centroid since the enhanced noise from the galaxy subtraction process yields a much brighter completeness limit compared to the region from $20\leq r\leq 470$ 
pixels (see Section \ref{Completeness}). Finally, we impose a bright magnitude limit of {\it F814W} = 26 mag based on the study of Abell 1689 ($z=0.18$) from 
\cite{Alamo13}, extrapolating to the redshift of J07173724.    
        
\subsection{Parallel field}
\label{Background}

In Section \ref{Observations} we described the MACS J0717.5+3745 parallel field that we use for estimating the number of background counts of starlike objects 
in order to 
statistically correct our catalogue of GC candidates. The image for object detection was constructed by summing the {\it F435W}, {\it F606W}, and {\it F814W} 
images using \textsc{iraf/imcombine} to generate a combined image. Unlike the galaxy field, there was no need to perform galaxy subtraction or median 
filtering. \textsc{ppp} was run on the combined parallel field image using the same parameters for object detection that were used for the galaxy frame. The 
\textsc{daophot} tasks \textsc{phot} and \textsc{allstar} were run in the same manner on the individual filter parallel field images as done for the galaxy field, 
using the master parallel field object position file created by \textsc{ppp}. PSF models were constructed for each filter of the parallel field following 
the same procedure as the galaxy field. The same aperture corrections applied to the galaxy images were also used for the parallel field frames to make a final 
object catalogue of the background field population. 
 
\subsection{Completeness function}
\label{Completeness}
Since the ability to detect faint objects is a function of magnitude, we need to map out the completeness function in order to make the necessary corrections to 
compensate for `missing' objects at the faint end, and thus determine the magnitude limit of our data. We followed the standard procedure of adding artificial 
stars of various magnitudes to our images, and followed the identical steps for object detection, classification, and photometric measurements that were 
performed on the galaxy and parallel fields. For the galaxy field, we used \textsc{daophot/addstar} to add 250 artificial stars per image per filter, for a 
total of 140 artificial frames (35,000 stars in total), having randomly assigned magnitudes in the range $25<{\it F814W}<32$. Artificial stars were positioned at 
random locations on the galaxy-subtracted images, excluding regions containing bright objects. The artificial stars were generated using the PSF models 
constructed for the galaxy and parallel fields. The magnitudes of the artificial stars in the {\it F435W} and {\it F606W} images were based on the average 
colour of GC candidates measured from the object catalogue for J07173724 (${\it F606W}-{\it F814W}=0.6$ and ${\it F435W}-{\it F814W}=1.5$). For a given trial, 
colour offsets were applied to the {\it F814W} magnitudes and then the three filter images were combined. The galaxy image was median filtered in the exact 
same manner as the original galaxy image. Object detection using \textsc{PPP} and \textsc{daophot} photometry measurements were done using the same settings 
and procedures as that used for the original galaxy frame. The appropriate aperture corrections were applied, and the input/output measured magnitudes were 
compared after culling objects based on the {\it sharp} and {\it chi} shape parameters from \textsc{allstar}.
 
\begin{figure}
\includegraphics[width=\columnwidth]{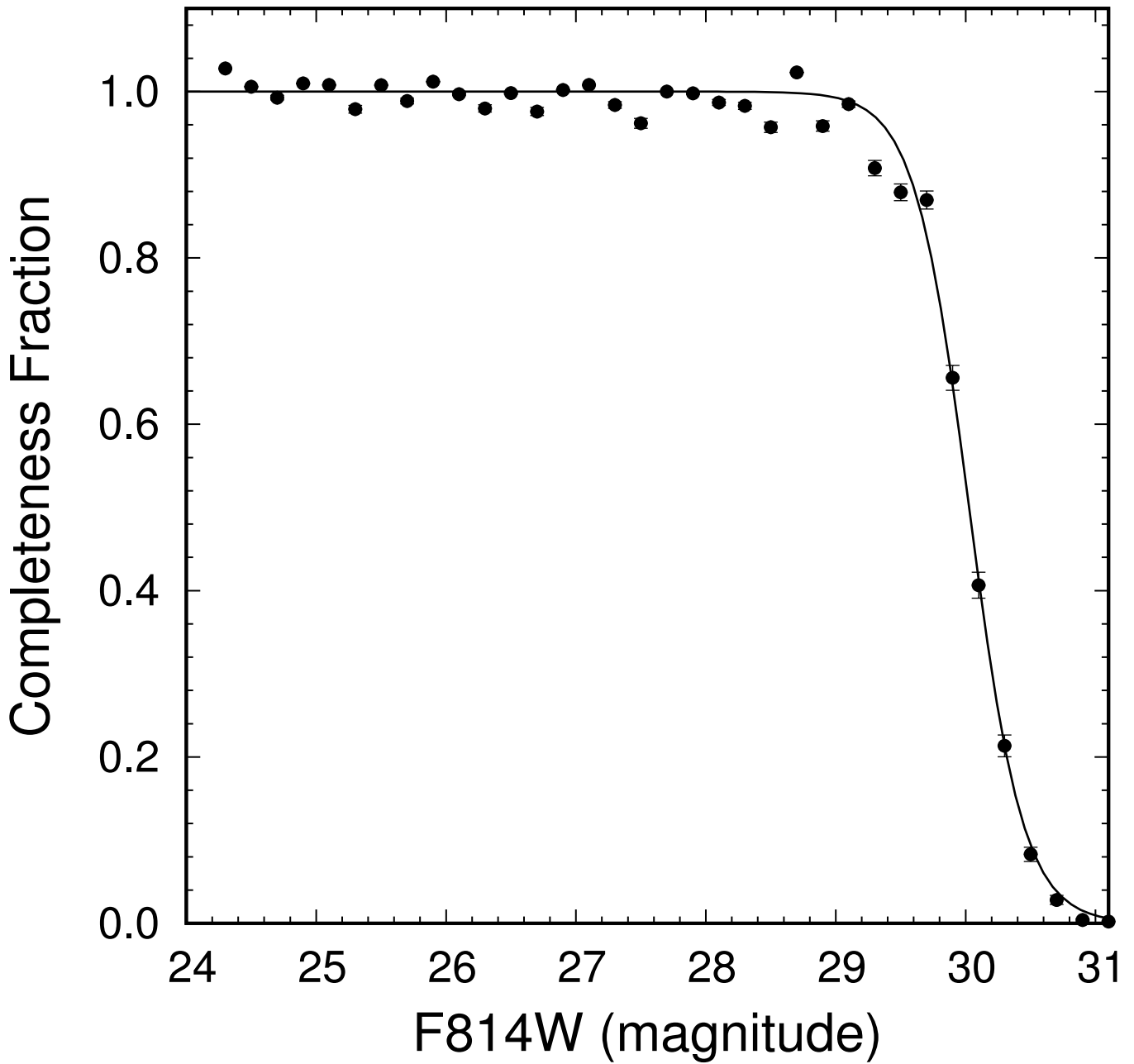}
\caption{{\it F814W} photometric completeness function for the J07173724 galaxy field. The solid curve depicts a sigmoid function with $\alpha=4.78\pm 0.23$ 
and a 50 per cent completeness fraction at $m_{0}=30.03\pm 0.01$ mag.}
\label{GalaxyCompletenessFunctionF814W}
\end{figure}

The completeness function, the fraction of recovered artificial stars per input magnitude bin, for the {\it F814W} galaxy field is shown in 
Fig. \ref{GalaxyCompletenessFunctionF814W}. The uncertainty for each mag bin is calculated assuming Poisson and Binomial statistics, and is given by the 
variance $\sigma^{2}=f(1-f)/n_{add}$, where $f=n_{rec}/n_{add}$ \citep{Bolte89}. Here $n_{add}$ is the number of stars added per magnitude bin, 
$n_{rec}$ is the number of recovered stars, and $f$ is the completeness fraction. 

The faint magnitude limit of the data is defined as the magnitude where the completeness fraction reaches 50 per cent \citep{Harris90}. This mag limit was 
estimated based on fitting the sigmoid-type function $f(m)=\{1+\exp{[\alpha(m-m_{0})}]\}^{-1}$, where $m_{0}$ is the magnitude at which $f=50$ per cent, 
and $\alpha$ depends on the steepness of the completeness function \citep[see eq. 2 from][]{Harris16}. A non-linear least-squares fit to the {\it F814W} 
galaxy field completeness function yields $\alpha=4.78\pm 0.23$ and $m_{0}=30.03\pm 0.01$ mag (solid curve in Fig. \ref{GalaxyCompletenessFunctionF814W}).    

\begin{figure}
\includegraphics[width=\columnwidth]{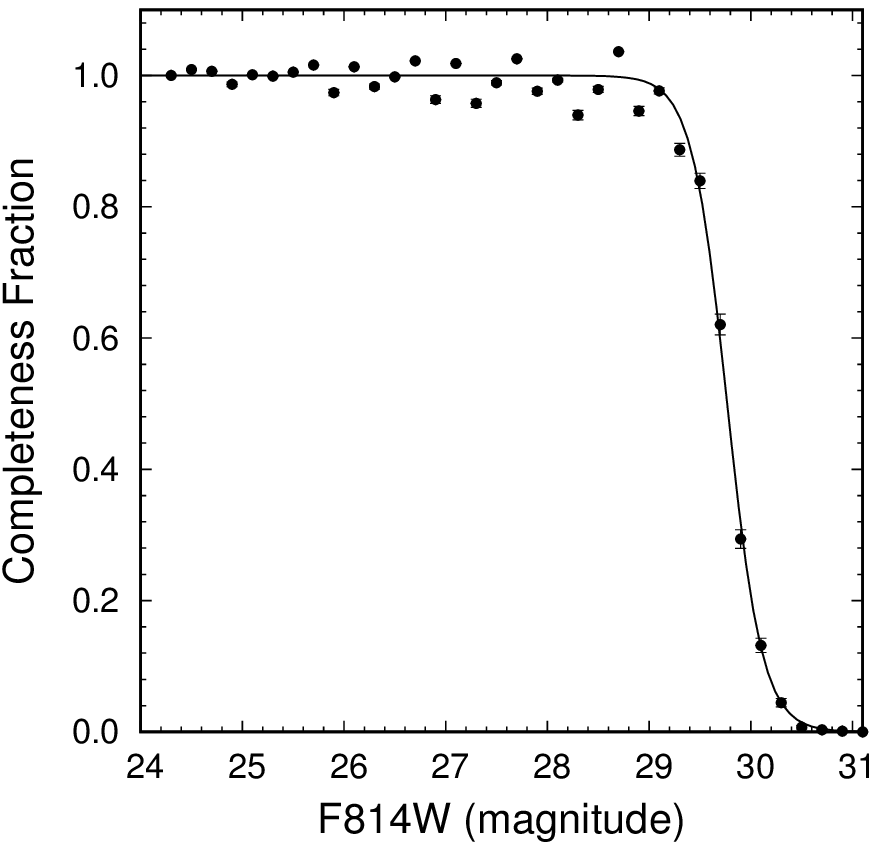}
\caption{Completeness function for the parallel field ({\it F814W}). The best-fitting sigmoid function, indicated by the solid curve, is characterised by 
$\alpha=5.77\pm 0.32$ and a 50 per cent completeness fraction at $m_{0}=29.77\pm 0.01$ mag.}
\label{ParallelFieldCompletenessFunctionF814W}
\end{figure}

Adopting the galaxy field procedure to the parallel field, we find $\alpha=5.77\pm 0.32$ and $m_{0}=29.77\pm 0.01$ mag ({\it F814W}). Since the 50 per cent 
magnitude limit is brighter for the parallel field, we impose a faint magnitude cutoff of {\it F814W} = 29.8 mag for both the galaxy and parallel field catalogues 
(see Fig. \ref{ParallelFieldCompletenessFunctionF814W}). 

For the magnitude range $26.0\leq {\it F814W}\leq 29.8$ mag, the standard deviation of the GC magnitude errors increase from a median of 
$\sigma_{\it F814W}=0.03$ at the bright-end to $\sigma_{\it F814W}=0.13$ at the faint-end cutoff. For the parallel field, $\sigma_{\it F814W}$ varied from a median 
of $0.02$ at ${\it F814W}=26$ mag to $\sigma_{\it F814W}=0.15$ at the faint limit. Systematic errors in the photometry were estimated based on the average magnitude 
difference of known input and measured artificial star magnitudes \citep{Bridges91}. For the galaxy field, we find a mean systematic error in the 
$26\leq {\it F814W}\leq 29.8$ mag range of +0.01 mag. Using the same filter and magnitude range, the mean systematic error for the parallel field was 
found to be -0.02 mag. Since the systematic errors for the galaxy and parallel fields are small, no corrections were applied to the GC magnitudes. 

\begin{figure}
\includegraphics[width=\columnwidth]{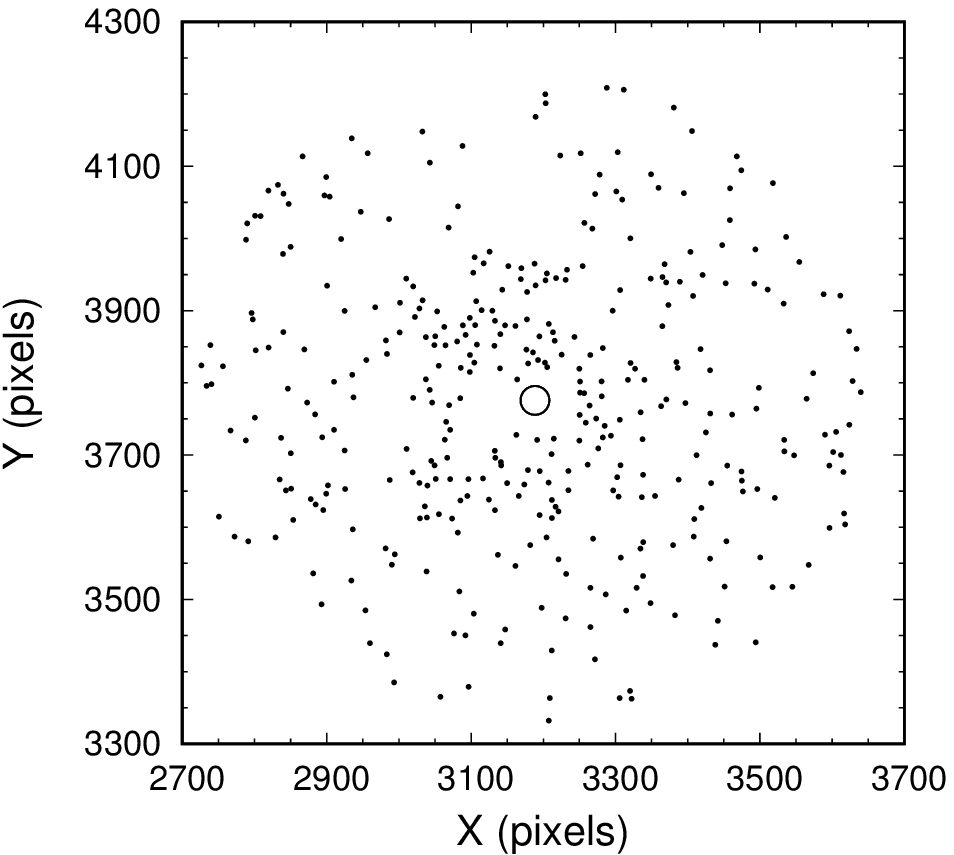}
\caption{Spatial distribution of 365 globular cluster candidates surrounding J07173724. The solid circle near the centre shows the inner 20 pixel radius 
masked out region.}
\label{SpatialPositions}
\end{figure}

To summarise, we select GCs by applying shape ($-1< {\it sharp} < +1$ and ${\it chi} < 4$), radius from galaxy centroid ($20\leq r\leq 470$ pixels), and magnitude 
($26\leq {\it F814W}\leq 29.8$) criteria to the object catalogue, obtaining a total of 365 GC candidates. The spatial distribution of our GC candidates 
is displayed in Fig. \ref{SpatialPositions}.

\section{Results}
\subsection{Colour distribution}

\begin{figure}
\includegraphics[width=\columnwidth]{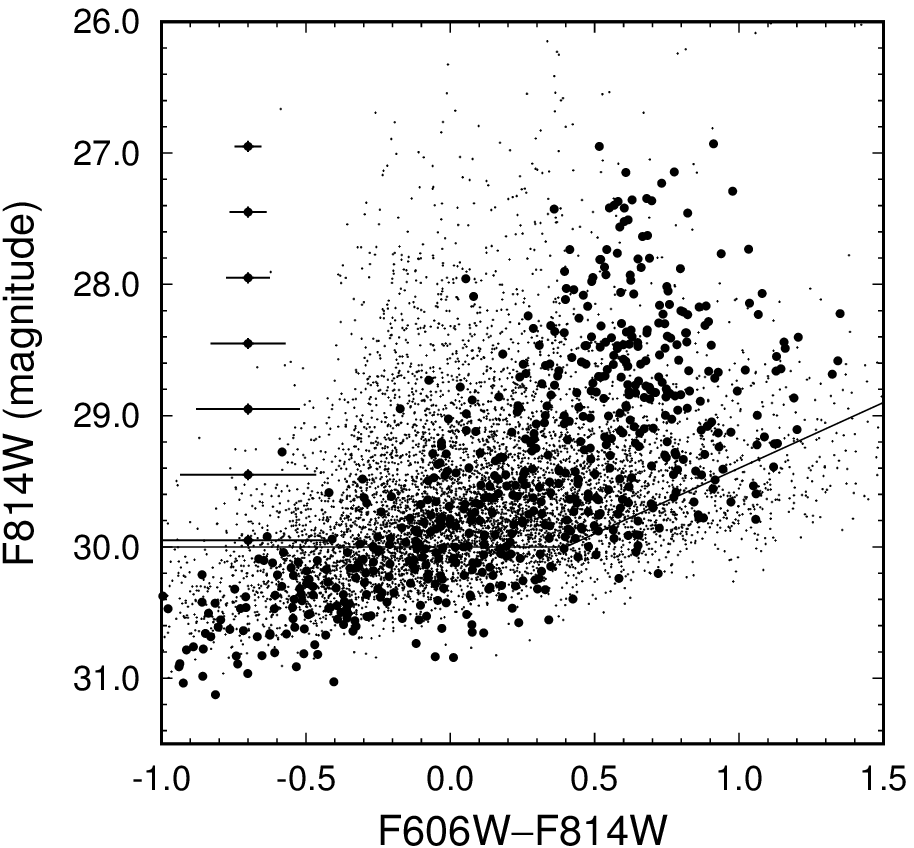}
\caption{The colour-magnitude ($F606W-F814W$ vs. $F814W$) diagram of objects selected from the galaxy and parallel fields using identical magnitude and 
shape criteria. The solid circles depict the GC candidates from the galaxy field and the dots represent objects selected from the much larger parallel field. 
The average colour uncertainties for the J07173724 field are given by the error bars. The 50 per cent completeness limit for the $F606W-F814W$ colour is 
indicated by the solid lines.}
\label{CMDgalaxyfield}
\end{figure}

For most galaxies the colour distribution of GCs displays a bimodal characteristic with the presence of a blue and red subpopulation. The ability to resolve 
multimodal colour distributions is dependent upon the wavelength separation of the filters used to observe the GC population and the precision of photometric 
measurements. In Fig. \ref{CMDgalaxyfield} we compare the $F606W-F814W$ vs. $F814W$ colour-magnitude distribution of GC candidates surrounding J07173724 with 
objects culled from the parallel field using the same selection criteria as the galaxy field. The solid circles represent GCs that have 
$F814W\geq 26$ mag, $-1< {\it sharp} < +1$, ${\it chi} < 4$, and radii $20\leq r\leq 470$ pixels from the photometric centroid of J07173724. The average 
uncertainty of the $F606W-F814W$ colour as a function of $F814W$ magnitude for the GCs is depicted by the errors bars. The 50 per cent completeness limit in 
$F606W-F814W$ colour of the galaxy field is shown by the solid lines. Objects selected from the parallel field are indicated with dots, 
and are extracted from an area that is $65\times$ larger than the galaxy field. Visual inspection of Fig. \ref{CMDgalaxyfield} shows that the colour-magnitude 
distribution of objects from the galaxy and parallel fields are different. GCs show a preferred colour of $F606W-F814W\sim 0.6$, while the parallel field objects 
cluster near  $F606W-F814W\sim 0.0$. 

\subsubsection{Testing for bimodality}

\begin{figure}
\includegraphics[width=\columnwidth]{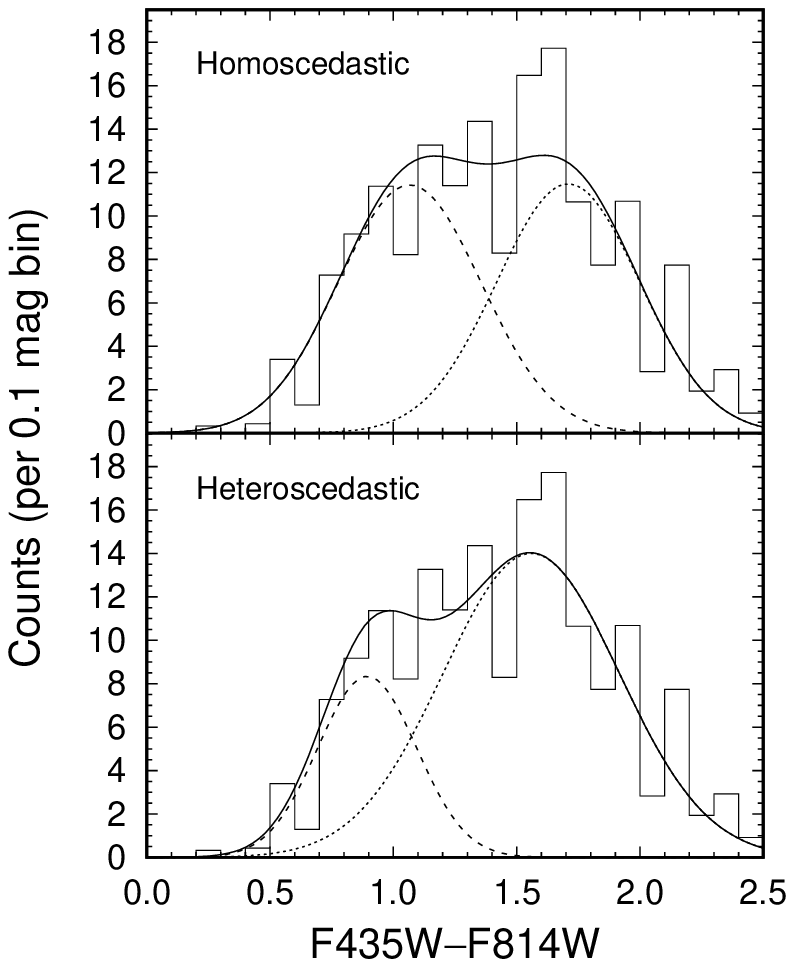}
\caption{$F435W-F814W$ colour histogram for background-corrected GC candidates with $26\leq F814W\leq 29$ in bins of $0.1$ mag. Only GCs with $20\leq r\leq 470$ pixels 
from the galaxy centroid are included. The best-fitting Gaussian for the blue peak (dashed line), red peak (dotted line), and the combined sum of blue and red (solid curve) 
is given for the homoscedastic (top panel) and heteroscedastic (bottom panel) case.}

\label{ColourHisF435F814}
\end{figure}

\begin{figure}
\includegraphics[width=\columnwidth]{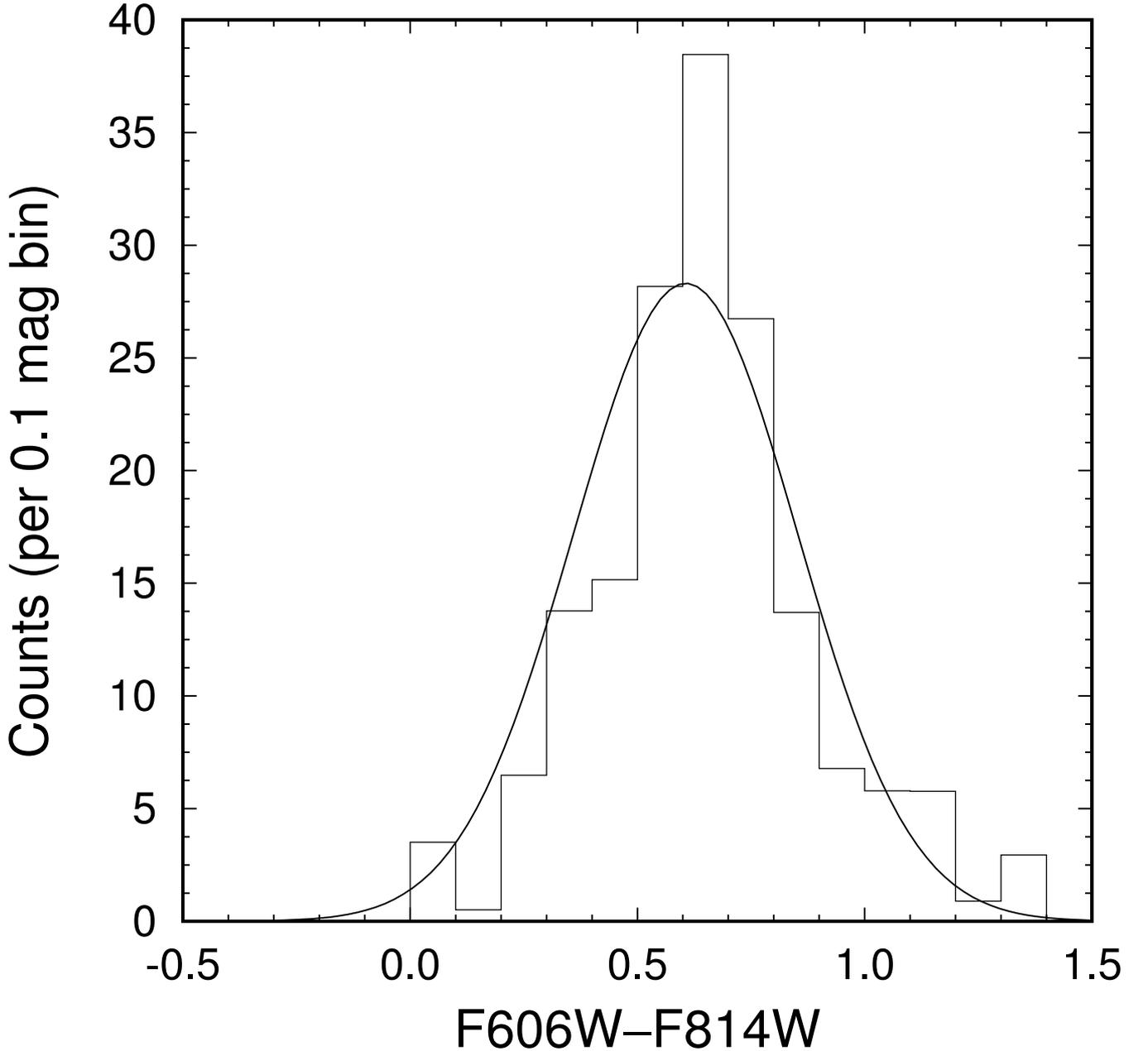}
\caption{Colour distribution for background-corrected GCs selected using the same $F814W$ magnitude and radial constraints as that used for the $F435W-F814W$ 
histogram. The solid curve is a best-fitting Gaussian with the peak and width given by \textsc{GMM} ($\mu=0.606$ and $\sigma=0.247$).} 
\label{ColourHisF606F814}
\end{figure}

As a test for bimodality in the colour distribution of our GC sample, we construct colour histograms in 0.1 mag bins using $F435W-F814W$ and $F606W-F814W$ 
colours (Fig. \ref{ColourHisF435F814} and Fig. \ref{ColourHisF606F814}). To ensure that the data are 100 per cent complete, we select GCs that have magnitudes 
in the range of $26\leq {\it F814W}\leq 29$ (179 GC candidates). Using the transformation of $F814W$ to the Johnson $V$-band (given by Alamo-Mart\'{i}nez et al. 
2013) and our adopted distance modulus, $F814W=26$ mag is equivalent to an absolute magnitude of $M_{V}=-12.5$ mag. The colour histograms are 
background-corrected using objects from the parallel field that have been selected using the same criteria as applied to the galaxy field. The number of parallel 
field objects have been corrected to compensate for the difference in area between the parallel field region and the area used to select GCs from the galaxy field. 
Since the parallel field is $\sim 65\times$ larger than the galaxy field, the parallel field counts are multiplied by an area correction factor of 0.0153 before 
subtraction from the galaxy GC counts. The total number of background-corrected GC candidates is 166, compared to 179 objects prior to applying the background correction 
measured from the parallel field. The uncertainties in the colour increase to $\Delta (F435W-F814W)=0.28$ mag and $\Delta (F606W-F814W)=0.17$ mag in the 
$F814W=28.9-29.0$ mag bin.      

To search for possible unimodal and bimodal peaks in the colour distributions, we use the Gaussian Mixture Modelling software (\textsc{GMM}) from 
\citet{Muratov10} to quantify the maximum likelihood that the colour peaks are best described by one or two modes. The \textsc{GMM} algorithm assumes that each 
population mode can be best described using a Gaussian function. This allows the maximum likelihood function to be fully characterised mathematically. We use the 
\textsc{GMM} software to test whether a unimodal, bimodal with different Gaussian variances (heteroscedastic), or bimodal with the same variance (homoscedastic), 
provides a better fit to the $F435W-F814W$ and $F606W-F814W$ colour histograms.  

For the $F435W-F814W$ colour histogram, a unimodal fit yields a peak of $\mu=1.399\pm 0.035$ with a Gaussian variance of $\sigma= 0.433\pm 0.020$, where the errors 
are calculated using a non-parametric bootstrap (\textsc{GMM} fit parameters are summarised in Table 1). A heteroscedastic bimodal fit gives $\mu_{1}=0.894\pm 0.146$, 
$\sigma_{1}=0.200\pm 0.072$, $\mu_{2}=1.559\pm 0.166$, and $\sigma_{2}=0.357\pm 0.067$. A comparison of the unimodal distribution with the heteroscedastic bimodal case 
yields a parametric bootstrap probability that the unimodal case is rejected at the 92\% level. 

For the homoscedastic bimodal distribution, \textsc{GMM} gives $\mu_{1}=1.066\pm 0.067$, $\mu_{2}=1.709\pm 0.058$, and $\sigma_{1}=\sigma_{2}=0.290\pm 0.025$. A comparison 
of the homoscedastic case with the unimodal distribution indicates that the unimodal distribution is rejected at the 98.5\% level.   

The significance of the separation of the colour peaks is described by the statistic $D=|\mu_{1}-\mu_{2}|/[(\sigma_{1}^{2}+\sigma_{2}^{2})/2]^{0.5}$, which 
is a measure of the peak separation relative to the width of the variances \citep{Muratov10}. A value of $D>2$ signifies that a bimodal distribution is a better 
description of the colour distribution than a unimodal peak \citep{Ashman94,Muratov10}. For the heteroscedastic case, the \textsc{GMM} algorithm yields $D=2.30\pm 0.31$, 
while the homoscedastic colour split gives $D=2.30\pm 0.31$. Both cases indicate that a bimodal distribution is statistically a better fit than a single peak distribution.

The fraction of blue GCs was calculated based on the \textsc{GMM} probability of each cluster belonging to either the blue or red subpopulation. For the heteroscedastic 
bimodal case, 40 of the 166 background-corrected GCs are associated with the blue peak and 126 GCs with the red population. The blue fraction, defined as 
$f_{b}={N(blue)}/{N(total)}$, was found to be $f_{b}=0.241$. For the homoscedastic case, 80 clusters are associated with the blue mode and 86 with the red GCs, yielding 
$f_{b}=0.482$. Statistically, we are not able to ascertain whether the homoscedastic or heteroscedastic case provides a better bimodal fit to the GC colour distribution.  
However, the fit parameters from \textsc{GMM} indicates that the red GCs are $1-3\times$ more abundant than the blue GCs.  

The $F606W-F814W$ colour distribution histogram in Fig. \ref{ColourHisF606F814} appears unimodal from visual inspection. In fact, \textsc{GMM} gives a unimodal peak 
at $\mu=0.606\pm 0.021$ with $\sigma=0.247\pm 0.016$. A bimodal fit for the heteroscedastic case yields $\mu_{1}=0.592\pm 0.086$, $\sigma_{1}=0.089\pm 0.062$, 
$\mu_{2}=0.611\pm 0.139$, $\sigma_{2}=0.285\pm 0.078$, and $D=0.09\pm 1.00$. For the homoscedastic bimodal distribution we find $\mu_{1}=0.589\pm 0.032$, 
$\sigma_{1}=0.240\pm 0.024$, $\mu_{2}=0.620\pm 0.099$, $\sigma_{2}=0.240\pm 0.024$, and $D=0.09\pm 0.63$. The narrow separation of the bimodal colour peaks and the $D$ values 
indicates that the $F606W-F814W$ colour distribution is not bimodal. Comparing Fig. \ref{ColourHisF435F814} and Fig. \ref{ColourHisF606F814} shows that the colour 
separation between the $F606W$ and $F814W$ filters is inadequate to resolve the bimodality displayed in Fig. \ref{ColourHisF435F814}.

\begin{figure}
\includegraphics[width=\columnwidth]{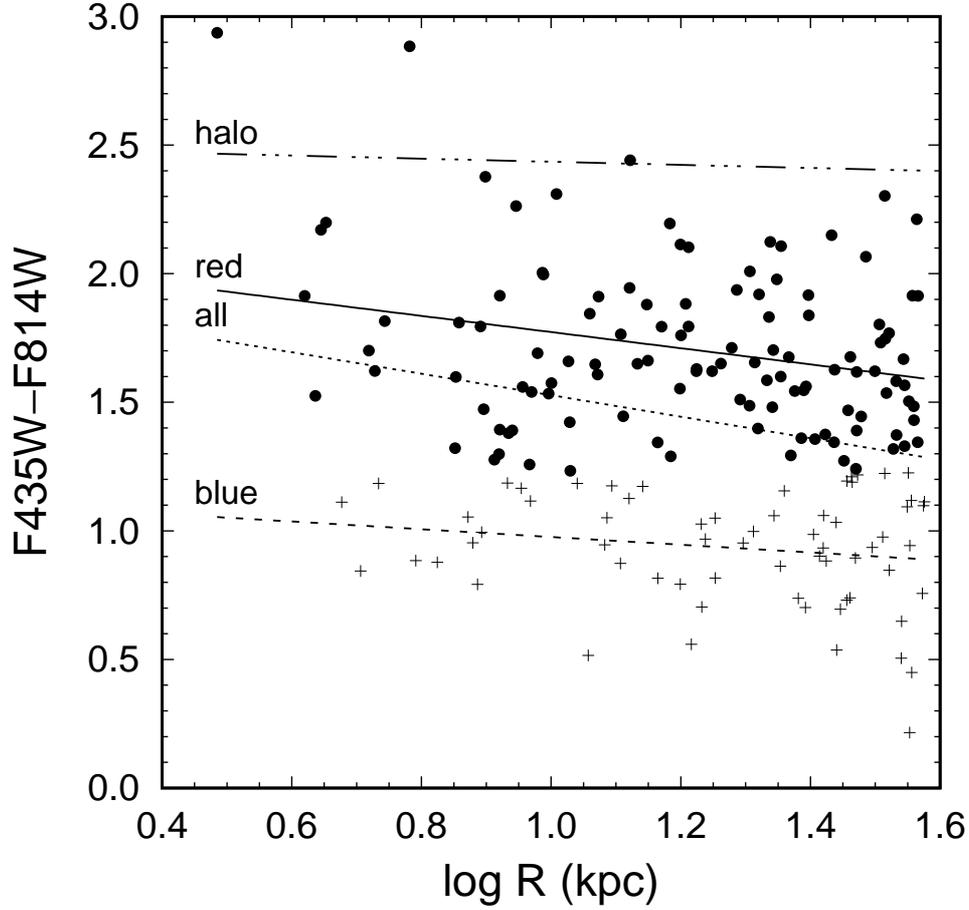}
\caption{Radial distribution of red and blue GC subpopulations. The solid points depict the red sample and the plus symbols indicate the position of the blue 
GCs. The solid line is a non-linear least-squares fit to the red GCs, the long dashed line is a fit to the blue sample, and the short dashed line is a fit to the 
combined blue+red population. For comparison, we include the colour profile of the galaxy halo (dashed-dotted line).}
\label{ColourRadiusF435F814}
\end{figure}

\begin{table}
\caption{The \textsc{GMM} fit parameters for the $F435W-F814W$ and $F606W-F814W$ colour histograms.}
\label{Table1}
\begin{tabular}{llllllll}
\hline
Colour & $\mu_{1}$ & $\mu_{2}$ & $\sigma_{1}$ & $\sigma_{2}$ & $D$ & Blue Fraction & Mode \\
\hline
$F435W-F814W$ & $1.399\pm 0.035$ &  & $0.433\pm 0.020$ &  &   &  & Unimodal\\
 & $0.894\pm 0.146$ & $1.559\pm 0.166$ & $0.200\pm 0.072$ & $0.357\pm 0.067$ & $2.30\pm 0.31$ & $0.241$ & Bimodal (heteroscedastic)\\
 & $1.066\pm 0.067$ & $1.709\pm 0.058$ & $0.290\pm 0.025$ & $0.290\pm 0.025$ & $2.30\pm 0.36$ & $0.482$ & Bimodal (homoscedastic)\\ 
$F606W-F814W$ & $0.606\pm 0.021$ &  & $0.247\pm 0.016$ &  &  &  & Unimodal\\
& $0.592\pm 0.086$ & $0.611\pm 0.139$ & $0.089\pm 0.062$ & $0.285\pm 0.078$ & $0.09\pm 1.00$ & 0.248 & Bimodal (heteroscedastic)\\
& $0.589\pm 0.032$ & $0.620\pm 0.099$ & $0.240\pm 0.024$ & $0.240\pm 0.024$ & $0.09\pm 0.63$ & $0.479$ & Bimodal (homoscedastic)\\ 
\hline
\end{tabular}
\end{table}

\subsubsection{Radial colour gradient}

The average colour of GCs has been shown in many studies to become bluer with increasing radius from the host galaxy 
\citep[e.g.][]{Harris09,Liu11,Blom12,Harris16}. To look for this trend for J07173724, we used the \textsc{GMM} heteroscedastic and homoscedastic 
bimodal fits of the $F435W-F814W$ colour histogram. Based on the midpoint of the colour peaks ($F435W-F814W=1.226/1.388$ for the heteroscedastic/homoscedastic 
case), we divide our GC sample into red and blue subpopulations.

In Fig. \ref{ColourRadiusF435F814} we show the $F435W-F814W$ radial distribution of red (solid points) and blue (plus symbols) GCs that have $26\leq F814W\leq 29$ 
and are $20\leq r\leq 470$ pixels from the galaxy centroid for the heteroscedastic case. The conversion of length scale from pixel to kpc has been done using a pixel scale of 
$0.03~\mbox{arcsec}~\mbox{pixel}^{-1}$, along with our adopted distance modulus of $(m-M)=39.34$. A linear least-squares fit to the red sample (solid line) 
yields $F435W-F814W=(-0.314\pm 0.111)R+(2.087\pm 0.138)$, while a best-fitting line to the blue population (long dashed line) gives 
$F435W-F814W=(-0.150\pm 0.106)R+(1.127\pm 0.138)$. For the homoscedastic case, the gradients are very similar to the previous case with 
$F435W-F814W=(-0.308\pm 0.112)R+(2.153\pm 0.137)$ for the red sample and $F435W-F814W=(-0.115\pm 0.108)R+(1.167\pm 0.141)$ for the blue GCs. A fit to the combined 
blue+red population yields $F435W-F814W=(-0.418\pm 0.131)R+(1.945\pm 0.165)$ (short dashed line in Fig. \ref{ColourRadiusF435F814}). The rms dispersion of each 
subpopulation about the best fitting line is found to be $\sigma_{rms}^{red}=0.31$ and $\sigma_{rms}^{blue}=0.21$ for the heteroscedastic case, and $\sigma_{rms}^{red}=0.29$ 
and $\sigma_{rms}^{blue}=0.25$ for the homoscedastic distribution. For the combined blue+red population we find $\sigma_{rms}^{all}=0.46$.

Both the blue and red GCs show a small negative slope with radius, with the red population having a slightly steeper negative gradient. The combined blue+red GCs 
have a larger gradient than either blue or red sample. These results are consistent with most studies \citep[e.g.][]{Liu11} and is 
normally attributed to an increase in the fraction of blue GCs with increasing distance from the host galaxy. In Fig. \ref{ColourRadiusF435F814} we also show the colour profile of 
the host galaxy halo (dashed-dotted line) having $F435W-F814W\sim 2.5$.

\subsection{Background field contamination}
\label{Backcontamination}

As stated in Section \ref{Background}, we correct our catalogue of GC candidates for background starlike objects by using the MACS J0717.5+3745 
parallel field. Using the same procedure for the detection and classification of objects from the parallel field as was used for the galaxy field, 
we can statistically correct for fore/background objects by subtracting the parallel field object counts from the galaxy field counts. Since the 
parallel field is $\sim 65\times$ larger in area than the region used to select GC candidates from the galaxy field, the parallel field object 
counts are normalised with respect to area (i.e. multiplied by a factor of 0.0153; see Section 3.1). 

\begin{figure}
\includegraphics[width=\columnwidth]{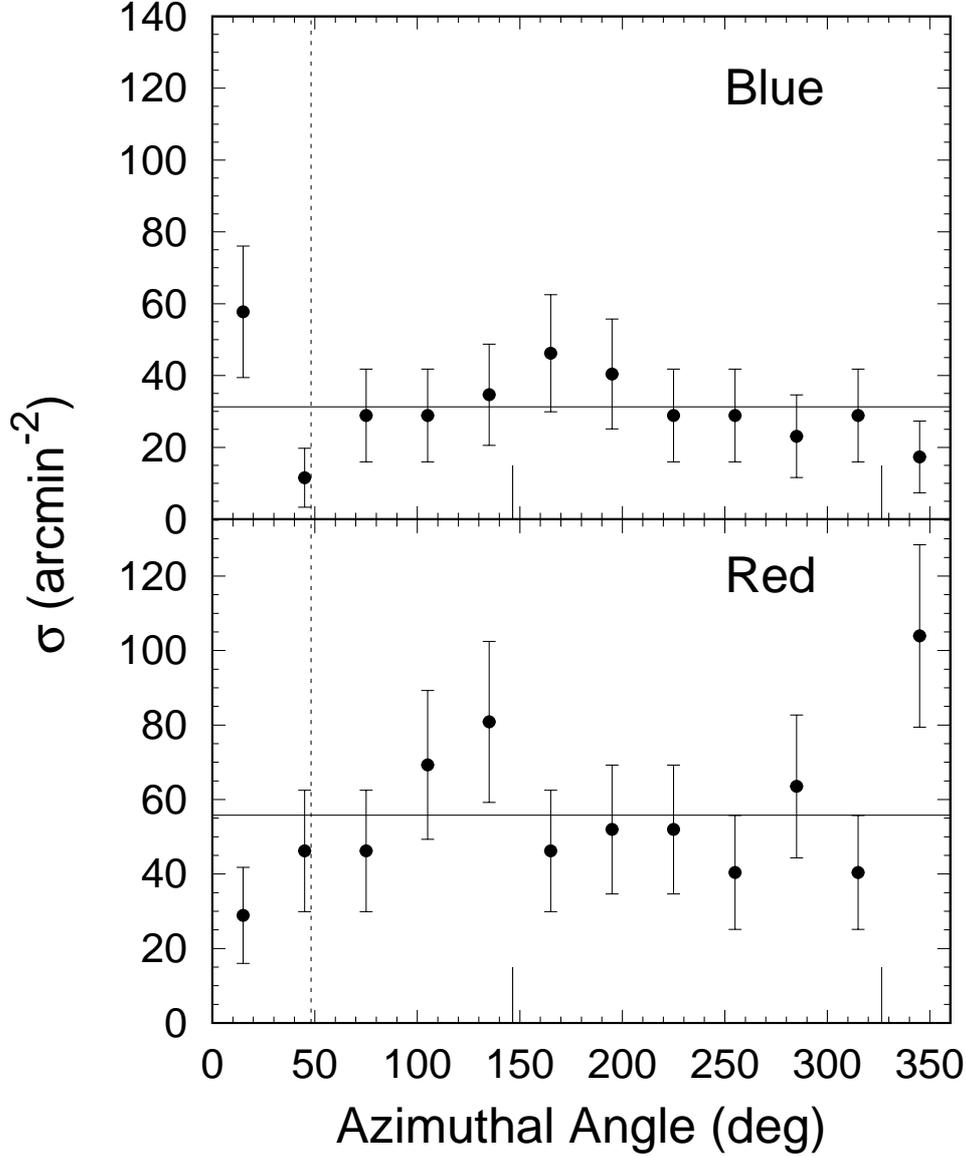}
\caption{The azimuthal density distribution of blue and red GC candidates. Objects are binned into $30\degr$ sectors with the mean density of each sample given 
by the horizontal line. The short vertical solid lines indicate the orientation of the isophotal major axis of J07173724. The dashed vertical line is the position 
angle of the radial direction to the centre of the background galaxy cluster MACS J0717.5+3745.}
\label{azimuthal}
\end{figure}

To determine if the fore/background counts are expected to have a significant impact on the results of this study, we show in Fig. \ref{azimuthal} 
the azimuthal density distribution of blue and red GC candidates surrounding J07173724 (using a colour divide of $F435W-F814W=1.226$). The isophotal major 
axis of the galaxy is indicated by the 
two short solid vertical lines ($146\degr$ and $326\degr$), where azimuthal angles are measured in the positive direction from east to north. The vertical 
dashed line represents the position angle from the centroid of J07173724 to the centre of the background galaxy cluster MACS J0717.5+3745 
\citep[$z=0.545$;][]{Jauzac17}. If our GC candidates are significantly affected by fore/background contamination, we would not expect that the density of GCs 
would be greatest along the major axis of the host galaxy, as is observed. In fact, many studies have shown that the spatial extent of the GC population is correlated 
with the position angle of the isophotal major axis of the host galaxy \citep[e.g.][]{Hargis14}.

If we assume that a large number of our GC candidates are background objects associated with the MACS J0717.5+3745 galaxy cluster, we would 
naively expect that the density of GC candidates would increase toward the direction of the cluster centre. From Fig. \ref{azimuthal} we see that this is not the case, 
and in fact the position angle towards the centre of MACS J0717.5+3745 is nearly orthogonal to the major axis of the isophote of J07173724, and that the density of GCs 
is a minimum toward the galaxy cluster centre. We thus expect that the contribution of fore/background objects to our sample of GC candidates, after statistically correcting 
for background starlike objects measured in the parallel field, will not significantly affect the results of our study.       
 
\subsection{Luminosity function}

\begin{figure}
\includegraphics[width=\columnwidth]{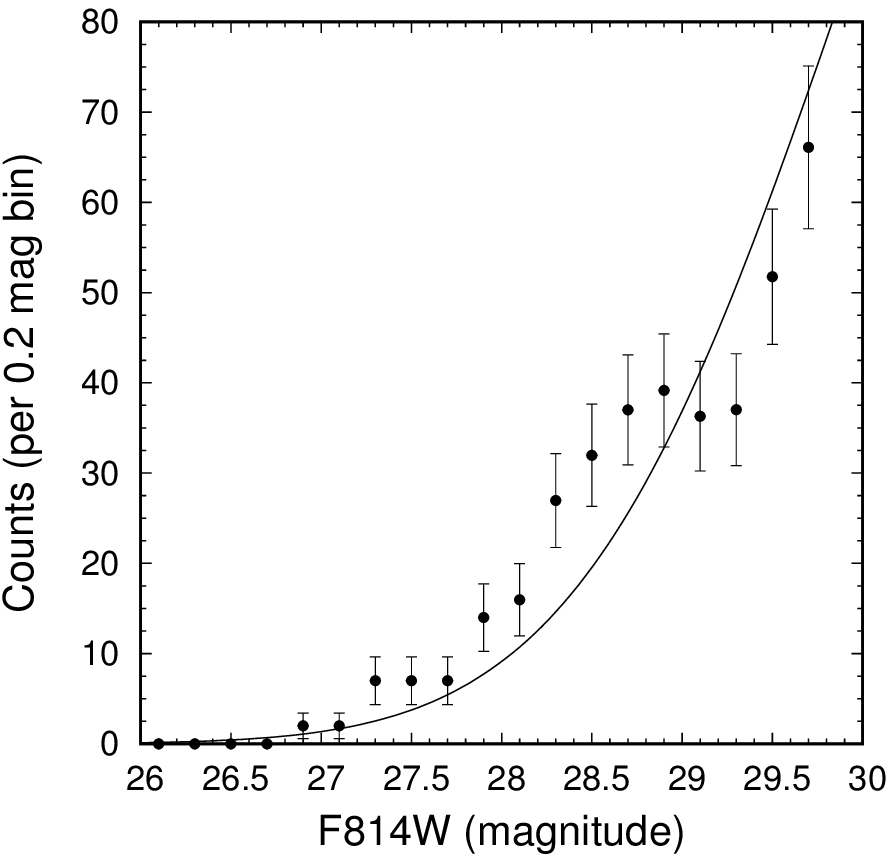}
\caption{Globular cluster luminosity function for J07173724. The GC counts have been corrected for magnitude completeness and contamination from the 
background field population. The best-fitting Gaussian function with $m_{TO}=31.2\pm 0.2$ mag, $\sigma=1.4\pm 0.1$, and $A=132.3\pm 11.1$ is depicted by the 
solid line.}
\label{LuminosityFunction}
\end{figure}

The globular cluster luminosity function (GCLF) is generally fit using a Gaussian function, $N=A\exp[-(m-m_{TO})/2\sigma^{2}]$ 
\citep[e.g.][]{Harris91,Jacoby92,Brodie06}, and is used to estimate the fraction of the GC population that is undetected due to the limiting magnitude of 
the data. The turnover in the GCLF for many galaxies has been found to be $M_{V}\sim -7.4$, with a Gaussian width of $\sigma\sim 1.4$ \citep[e.g.][]{Jordan07}. 

In Fig. \ref{LuminosityFunction} we plot the $F814W$ luminosity function for GCs associated with J07173724. The number of GCs per 0.2 magnitude bin has been 
corrected for completeness and background-subtracted for those GCs with $26.0\leq F814W\leq 29.8$, and with $20\leq r\leq 470$ pixels from the galaxy 
centroid (see Table \ref{Table2}). Since our observations do not reach the turnover in the Gaussian shape of the GCLF, we fix the turnover magnitude and 
Gaussian width, and fit for the amplitude. We have adopted a turnover magnitude of $M_{V}=-7.4\pm 0.2$ mag and a width of $\sigma=1.4\pm 0.1$. These values 
are consistent with those used in several studies of elliptical galaxies \citep[e.g.][]{Barkhouse01,Brodie06,Jordan07,Harris09,Alamo13}. 
Following \citet{Alamo13} we convert $M_{V}$ to $M_{F814W}$ and adopt a turnover absolute magnitude of $M_{F814W}=-8.1\pm 0.2$ mag. Using a distance modulus 
of $(m-M)=39.34$, our turnover apparent magnitude in the $F814W$ filter is $m_{TO}=31.2\pm 0.2$ mag. A non-linear least-squares fit to the GCLF with 
$m_{TO}=31.2\pm 0.2$ mag and $\sigma=1.4\pm 0.1$, yields an amplitude of $A=132.3\pm 11.1$, with $\chi^{2}_{\nu}=2.3$ (solid line in 
Fig. \ref{LuminosityFunction}).     

\begin{table}
\caption{Globular cluster counts corrected for completeness ($N_{glob}$), area-normalised background counts ($N_{bkg}$), and net 
background-subtracted GC counts ($N_{net}$) in 0.2 mag bins.}
\label{Table2}
\begin{tabular}{lccc}
\hline
$F814W$ & $N_{glob}$ & $N_{bkg}$ & $N_{net}$\\
\hline
26.9 & $2.0\pm 1.4$ & $0.0\pm 0.0$ & $2.0\pm 1.4$\\
27.1 & $2.0\pm 1.4$ & $0.0\pm 0.0$ & $2.0\pm 1.4$\\
27.3 & $7.0\pm 2.6$ & $0.0\pm 0.0$ & $7.0\pm 2.6$\\
27.5 & $7.0\pm 2.6$ & $0.0\pm 0.0$ & $7.0\pm 2.6$\\
27.7 & $7.0\pm 2.6$ & $0.0\pm 0.0$ & $7.0\pm 2.6$\\
27.9 & $14.0\pm 3.7$ & $0.0\pm 0.0$ & $14.0\pm 3.7$\\
28.1 & $16.0\pm 4.0$ & $0.0\pm 0.0$ & $16.0\pm 4.0$\\
28.3 & $27.0\pm 5.2$ & $0.0\pm 0.0$ & $27.0\pm 5.2$\\
28.5 & $32.0\pm 5.7$ & $0.0\pm 0.0$ & $32.0\pm 5.7$\\
28.7 & $37.1\pm 6.1$ & $0.1\pm 0.0$ & $37.0\pm 6.1$\\
28.9 & $39.2\pm 6.3$ & $0.0\pm 0.0$ & $39.2\pm 6.3$\\
29.1 & $36.4\pm 6.1$ & $0.1\pm 0.0$ & $36.3\pm 6.1$\\
29.3 & $37.1\pm 6.2$ & $0.1\pm 0.0$ & $37.0\pm 6.2$\\
29.5 & $51.9\pm 7.5$ & $0.1\pm 0.0$ & $51.8\pm 7.5$\\
29.7 & $66.5\pm 9.0$ & $0.4\pm 0.1$ & $66.1\pm 9.0$\\
\hline
\end{tabular}
\end{table}

\subsection{Radial profile}

\begin{figure}
\includegraphics[width=\columnwidth]{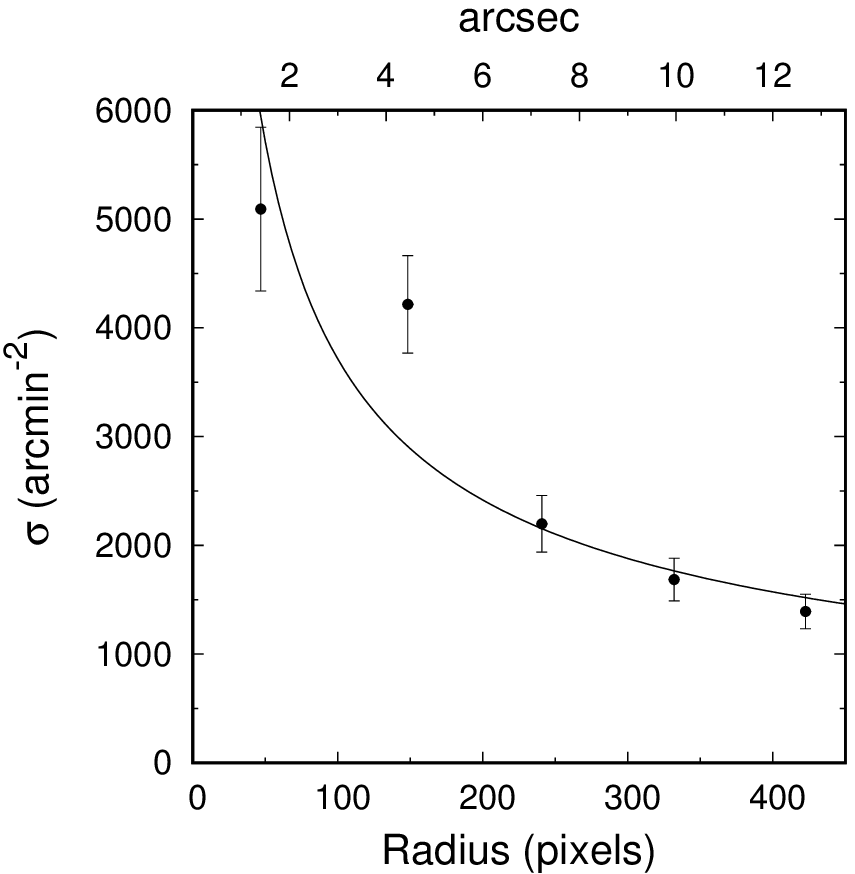}
\caption{Radial distribution of background-subtracted GCs corrected for completeness. The best-fitting powerlaw with $\alpha=-0.62\pm 0.13$ is indicated 
by the solid line.}
\label{RadialProfile}
\end{figure}

An estimate of the total number of GCs associated with J07173724 must take into consideration the radial limits in measuring the spatial distribution of 
objects. The radial distribution of the GC population was determined by dividing the $20\leq r\leq 470$ pixel region into five annuli, each 
having a width of 90 pixels. In  Fig. \ref{RadialProfile} we plot the radial spatial density ($\mbox{arcmin}^{-2}$) of completeness- and background-corrected 
GC counts. These data points are plotted using the geometric centre of each annuli given by $R=\sqrt{R_{in}R_{out}}$, where $R_{in}$ and $R_{out}$ represent 
the inner and outer radius boundary of each annuli, respectively. A non-linear least-squares fit of a powerlaw of the form $\sigma=AR^{\alpha}$ to the radial 
profile yields $A=64662\pm 45550$, $\alpha=-0.62\pm 0.13$, and $\chi^{2}_{\nu}=3.5$ (solid line in Fig. \ref{RadialProfile}).

\begin{figure}
\includegraphics[width=\columnwidth]{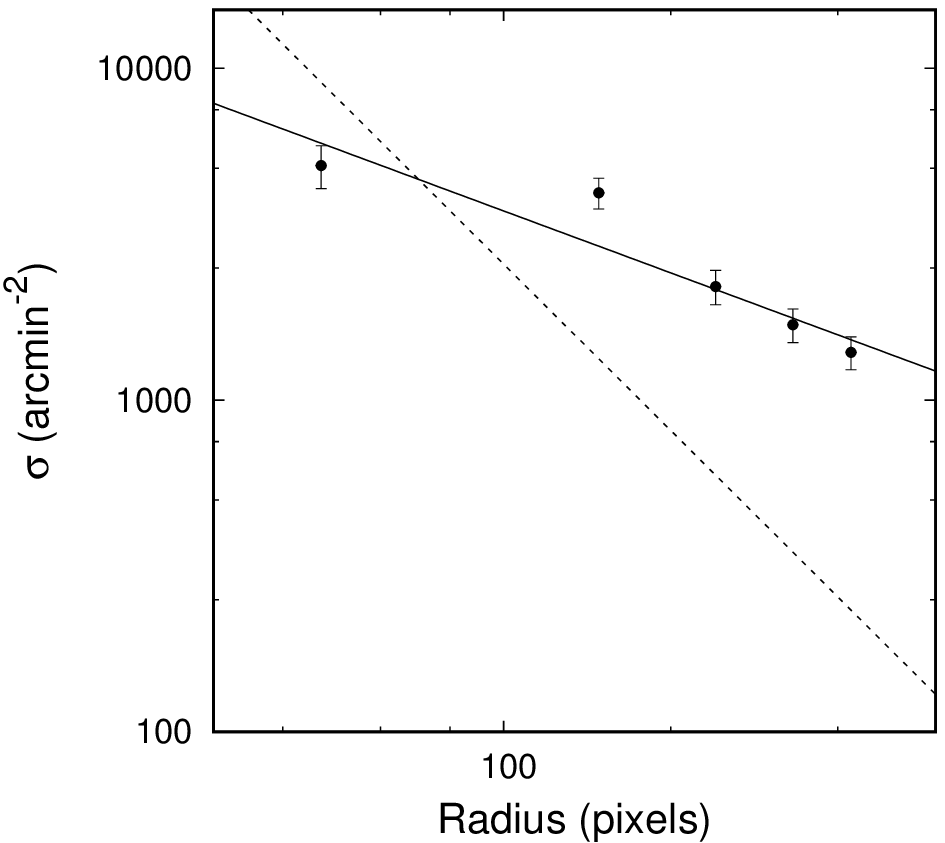}
\caption{Comparison of the GC radial distribution (solid line) with the $F814W$ surface brightness profile of the halo of J07173724 (dashed line). The 
surface brightness profile has been scaled vertically to aid the comparison. The GC profile has a powerlaw slope of $\alpha=-0.62$, while the surface 
brightness distribution of the halo has a slope of $\alpha\sim -1.6$.}
\label{SurfaceBrightness}
\end{figure}

The extent of the GC population is compared to the galaxy stellar halo in Fig. \ref{SurfaceBrightness}. The surface brightness profile for J07173724 
(dashed line) was measured from the $F814W$ image using the \textsc{ellipse} task in the \textsc{iraf/stsdas} package as described in Section 2.1, 
and shifted vertically to match the GC radial profile (solid line). To compare the slope of the surface brightness profile with the GC radial 
distribution, the radial dependence of the surface brightness was fit using a powerlaw. A non-linear least-squares fit to the galaxy halo yields 
$\mu\sim R^{-1.6}$, thus the GC population is spatially more extended than the halo light of the host galaxy. 

\subsection{Specific frequency}

As described in Section 1, the GC specific frequency, $S_{N}=N_{tot}10^{0.4(M_{V}+15)}$, is used as a statistic to quantify the number of GCs per unit galaxy 
luminosity \citep{Harris81}. The total number of GCs associated with J07173724, $N_{tot}$, can be estimated by first integrating over the radial density 
profile, and then correcting for incomplete coverage of the GC luminosity function. We adopted an inner radius limit of 1 kpc for integrating the powerlaw 
$\sigma=AR^{\alpha}$. This limit is based on several studies of GC populations of elliptical galaxies that show a dearth of GCs within 1 kpc of the host 
galaxy \citep{Harris91}. This has been attributed to dynamical friction and shock heating effects that dominate near the galaxy centre 
\citep{Weinberg93,Murali97,Gnedin99,Miocchi06,Brandt15}. For the outer radial limit, we use 50 kpc since, in general, most GC systems for ordinary non-cD 
ellipticals do not extend significantly past this radius \citep[e.g.][]{Hargis14}. Integrating the radial profile from 1 kpc (0.3 arcsec) to 
50 kpc (18.6 arcsec), we find  $N_{GCs}=523\pm 136$. If we select an outer radial limit of 25 (75) kpc, we find that $N_{GCs}\sim 200\,(918)$. Thus the estimated 
number of GCs is sensitive to our adopted outer radial limit.

To correct for magnitude incompleteness, we calculate the fraction of the GCLF (described by a Gaussian function) that is sampled by the observations and 
correct the number of GCs by scaling this number to compensate for the `missing' GCs. Taking the ratio of the number of GCs found by integrating the GCLF 
from $-\infty$ to the photometric limit of the data ($F814W=29.8$ mag), to the integral of the GCLF from $-\infty$ to $+\infty$, we find that $15.2\pm 4.9$ 
per cent of the GCLF is observed. Applying the radial and GCLF corrections, we estimate that the total number of GCs is $N_{tot}=3441\pm 1416$.

The calculation of specific frequency requires an estimate of $M_{V}$ for the host galaxy. We elected to use the {\it HST} ACS $F814W=16.30$ mag 
(extinction-corrected) value for J07173724 from the CLASH photometric catalogue \citep{Postman12} available at 
\url{https://archive.stsci.edu/missions/hlsp/clash/}. The transformation of $F814W$ to $M_{V}$ is given by $M_{V}=F814W-(m-M)+(V-I_{F814W})$, 
which is based on the conversion given by \citet{Alamo13}. Assuming that $V-I_{F814W}=0.83$ for an average elliptical galaxy at $z\sim 0.15$ (as adopted 
by Alamo-Mart\'{i}nez et al. 2013) and a distance modulus of $(m-M)=39.34$, we find $M_{V}=-22.21$. Using $N_{tot}=3441\pm 1416$ and $M_{V}=-22.21$, the specific 
frequency for J07173724 was found to be $S_{N}=4.5\pm 1.8$.

\section{Discussion}

Using the {\it HST} ACS Frontier Fields $F435W$, $F606W$, and $F814W$ observations of MACS J0717.5+3745, we have measured the properties of the globular 
cluster population associated with the foreground galaxy J07173724. Examination of the $F435W-F814W$ colour histogram has shown that a bimodal 
distribution provides a better description than a unimodal Gaussian function. The bimodal colour peaks at $\mu_{1}=0.894\pm 0.146$ and $\mu_{2}=1.559\pm 0.166$ for 
the heteroscedastic case, and $\mu_{1}=1.066\pm 0.067$ and $\mu_{2}=1.709\pm 0.058$ for the homoscedastic distribution. The division into separate colour peaks 
allows us to divide the GC candidates into a separate blue and red subpopulation. The separation of GCs into blue and red objects yields the interesting result 
that red GCs are $1-3\times$ more abundant than blue GCs, depending on the adopted bimodal fit. In comparison to other studies, galaxies on average have 
$\sim 4\times$ the number of blue compared to red GCs \citep{Harris16a}. Alternatively, a study of the GC population of NGC 1399 by \citet{Blakeslee12} using $g-I$ colour, 
finds that red GCs make up 70 per cent of the total population (compared to 76 per cent for J07173724 when adopting the heteroscedastic case). A recent study by 
\citet{Beasley18} describes the extreme case where NGC 1277 was found to have a unimodal distribution of red GCs. J07173724 is an important case since it 
contains a larger than average fraction of red GCs, and may be an example of a GC population that helps to bridge the gap between the majority of galaxies that 
are dominated by blue GCs \citep[e.g.][]{Jordan07} and NGC 1277. 

The difference in the ratio of the number of blue versus red GCs may provide an important clue to the galaxy formation process \citep[e.g.][]{Harris01}. In the 
monolithic collapse scenario for elliptical galaxy formation \citep{Eggen62}, one expects that the formation of GCs would result
in a unimodal colour distribution. The existence of blue and red subpopulations are hard to explain using the monolithic collapse model, especially since
the difference in GC colour appears to be due mainly to metallicity differences \citep[e.g.][]{Usher15}. An alternative formation mechanism is the model of
\citet{Searle78} where protogalactic gas clouds continue to fall into the galaxy over an extended period of time. The Eggen et al. model can be used to explain 
the `seed' galaxy along with the blue metal-poor GCs, while the red metal-rich GCs are formed from the protogalactic gas clouds that self-enrich with time. The
timescales involved with the accumulation of protogalactic fragments will have a direct impact on the number of episodes of GC formation, and whether
a bimodal or multimodal colour distribution is expected.

A GC formation model was proposed by \citet{Ashman92} and \citet{Zepf93} in which the merger of gas-rich disk galaxies, along with their metal-poor 
GCs, formed elliptical galaxies. The red GCs formed at a later time ($\sim 2$ Gyr) from enriched gas via mergers \citep{Ashman98}. A prediction of the 
merger model is that elliptical galaxies should have a bimodal colour distribution. Also expected is that red GCs will be more centrally 
concentrated than blue GCs. The merger model also suggests that red GCs should be more abundant than their blue counterparts. Our results for 
J07173724 may be consistent with the merger model in the sense that red GCs are more abundant than blue systems if we adopt the heteroscedastic bimodal case, 
and that the red GCs are more centrally concentrated. The main issue with the merger model involves providing an adequate explanation for the specific frequency 
of giant elliptical galaxies ($S_{N}=4.5$ for J07173724) from the merger of spiral galaxies with $S_{N}\sim 1$ \citep[e.g.][]{Harris01}. An additional constraint on 
the formation process of the GC population of J07173724 is that the galaxy halo is redder than the red GCs by $\Delta(F435W-F814W)\sim 0.5$ 
(see Fig. \ref{ColourRadiusF435F814}). 

The accretion model of \citet{Cote98,Cote00} posits that red GCs formed along with a seed galaxy (possibly a giant elliptical) that has a deep gravitational 
potential well, and thus able to hold onto the enriched gas expelled during the first generation of stellar evolution \citep{Carlberg84}. The blue GCs 
are accumulated from the accretion of low-mass galaxies which contain essentially metal poor GCs. Since the blue GC subpopulation is usually more extended 
than the red GC population, the accretion of blue GCs is expected to occur from dwarf galaxies at large radii. For example, \citet{Lim17} finds that blue GCs 
in NGC 474 are associated with shells and substructure, and thus consistent with the process of the accretion of low-mass galaxies at large radii.  
A potential problem with the accretion model is that for galaxies with a large fraction of blue GCs, the accretion of lots of blue GCs would also result in 
the accretion of a large number of metal-poor halo stars. This is expected to result in a halo colour that is bluer than the red GCs, in conflict with what is 
typically observed for GC systems, including J07173724 \citep[e.g.][]{Brodie06,Harris03,Strader04,Hargis14}.

The in situ model of \citet{Forbes97} suggests that metal poor GCs formed early in the history of the collapsing protogalaxy, and that red GCs, along 
with the majority of halo stars, formed $\sim 1-2$ Gyrs later out of enriched gas triggered by merger-induced starbursts. For J07173724, most of the 
halo stars would have formed after the red GCs in order to have an average $F435W-F814W$ colour redder by $\sim 0.5$ compared to the red GCs. 

The recent study of \citet{Beasley18} argues that the red-dominated GC population of NGC 1277 can be explained by an initial burst of star formation 
that forms the metal-rich red GCs, with little accretion of metal-poor blue GCs afterwards. This scenario represents an extreme case since typical GC 
populations of non-dwarf galaxies have a mixture of both blue and red GCs \citep{Jordan07}.  

The formation scenario that is consistent with the observed properties of the GC population of J07173724 is one which involves the combination of in situ and 
accretion models. The metal-rich red GCs are formed in situ from metal rich gas, followed shortly thereafter by the majority of the halo stars (giving the 
halo stars an average $F435W-F814W$ colour $\sim 0.5$ redder than the red GCs). The blue metal-poor GCs are accreted from low-mass dwarf galaxies, where the 
accretion process resulted in a ratio of 1:1 to as large as 3:1 for the number of red versus blue GCs. Since the majority of galaxies are more abundant in 
blue rather than red GCs, the accretion process has been extremely important in shaping the colour distribution of the GC population. This does not rule out 
extreme situations where accretion is unimportant compared to in situ formation, such as that of NGC 1277. For J07173724, a combination of in situ formation 
of red GCs and an accretion rate slower than average could explain the colour distribution of its GC population. 

\section{Conclusions}

We present the first measurement of the globular cluster population surrounding the elliptical galaxy J07173724. This galaxy is located in the foreground 
of the {\it HST} ACS observations of the galaxy cluster MACS J0717.5+3745. Based on {\it F435W} and {\it F814W} images, J07173724 is found to have 
a bimodal colour distribution with red GCs $1-3\times$ more abundant than blue GCs. The GC population is more extended than the halo light of 
the host galaxy, with the red GCs more concentrated toward the galaxy centroid than the blue clusters. The total number of GCs was estimated to 
be $N_{tot}=3441\pm 1416$, yielding a specific frequency of $S_{N}=4.5 \pm 1.8$. We conclude that our results are consistent with a mixed formation scenario 
in which the red GCs and halo stars are formed by an in situ process, while the blue GCs are acquired via the accretion of dwarf galaxies. 
   
\section*{Acknowledgements}

We thank the anonymous referee for comments that significantly improved this manuscript. We also thank Jose Diego for providing us with a mass estimate of J07173724 
from lens modeling of 
MACS J0717.5+3745. NC and WAB thank the University of North Dakota for financial support through the ND EPSCoR AURA program. We thank Harald Ebeling (IfA, University of Hawaii)
for providing the unpublished redshift of J07173724, and Tracy Clarke (U.S. Naval Research Laboratory) for useful discussions. Based on observations obtained with
the NASA/ESA Hubble Space Telescope, retrieved from the Mikulski Archive for Space Telescopes (MAST) at the Space Telescope Science Institute (STScI).
STScI is operated by the Association of Universities for Research in Astronomy, Inc. under NASA contract NAS 5-26555. This research has made use of the
NASA/IPAC Extragalactic Database (NED) which is operated by the Jet Propulsion Laboratory, California Institute of Technology, under contract with the National
Aeronautics and Space Administration.






\bsp	
\label{lastpage}
\end{document}